\documentstyle[aps,12pt]{revtex}
\newcommand{\bb}{\begin{equation}}
\newcommand{\ee}{\end{equation}}
\newcommand{\ba}{\begin{array}}
\newcommand{\ea}{\end{array}}

\begin {document}
\baselineskip 2.2pc

\title{Scattering of relativistic particles by a Coulomb field
       in two dimensions
        \thanks{published in Phys. Lett. A {\bf 260} (1999) 17-23.
        \copyright 1999 Elsevier Science B.V.}}
\author{Qiong-gui Lin\thanks{E-mail addresses:
        qg\_lin@163.net, stdp@zsu.edu.cn}}
\address{China Center of Advanced Science and Technology (World
	Laboratory),\\
        P.O.Box 8730, Beijing 100080, People's Republic of China\\
        and\\
        Department of Physics, Zhongshan University, Guangzhou
        510275,\\
        People's  Republic of China \thanks{Mailing address}}

\maketitle
\vfill

\begin{abstract}
{\normalsize The scattering of relativistic Dirac particles by a
Coulomb field $\pm Ze^2/r$ in two dimensions is studied and the
scattering amplitude is obtained as a partial wave series. For small
$Z$ the series can be summed up approximately to give a closed form.
The result, though being aproximate, exhibites some nonperturbative
feature and cannot be obtained from perturbative quantum
electrodynamics at the tree level.}
\end{abstract}
\vfill
\leftline {Keywords: relativistic scattering, Coulomb field, two
dimensions}
\leftline {PACS number(s): 03.65.Pm, 11.80.Et}
\newpage

\baselineskip 15pt %2.2pc
The scattering of particles by a Coulomb field is an important
subject in quantum mechanics as well as in classical mechanics,
because the Coulomb field describes real physical interations
between charged particles, say, electrons and nucleuses. In the
nonrelativistic case, exact results of the scattering amplitude
or the differential cross section in closed forms can be easily
obtained by solving the Schr\"odinger equation in parabolic
coordinates [1, 2]. This has been well known for a long   time.
For fast electrons (or  positrons) scattered by a nucleus, however,
the relativistic effect becomes significant and the Dirac equation
has to be employed. In this case the problem is rather complicated
and exact results can be obtained only in open forms, that is, in
partial wave series. For light nucleuses, the series can be summed up
approximately  to give a closed form for the differential cross
sectiion [3, 4]. This includes the
nonrelativistic result and a relativistic correction.
It is the same as the lowest-order result of quantum electrodynamics
(QED) [5]. Higher-order corrections involve more difficulty.
The problem has been investigated by many authors over several
decades [3-10]. (More references on the subject can be found in
Refs. [4] and [11].)

Mathematically one can consider the Coulomb scattering problem 
in general space dimension $d\ge 2$. It has been shown that the
(nonrelativistic) quantum-mechanical differential cross section
coincides with the classical one in $d=3$ only [12]. Among the
various space dimensions other than three, the two-dimensional
case seems physically relevant. In the recent years, there has been
wide interest in lower-dimensional field theories and condensed
matter physics. The Coulomb field may be of interest in some
two-dimensional physical models [13, 14] which may describe the
interaction of Chern-Simons vortex solitons [15-22]. The
nonrelativistic scattering by a Coulomb field in two dimensions
has been solved exactly [23]. The result exhibits new features
different from those  in three dimensions. For example, the
quantum-mechanical differential cross section is different from
the classical one as mentioned above, and the exact result is
different from the Born approximation. Thus the problem is of some
interest in itself as well. The purpose of this note is to extend
the  nonrelativistic result to the relativistic case.

Consider the scattering of a fast electron (position) with mass
$\mu$ and electric charge $-e$ ($e$) by a nucleus with charge $Ze$.
When the incident velocity is comparable with $c$, the speed of light
in vacuum, the process should be described by the relativistic Dirac
equation. We begin with the stationary Dirac equation of the electron
(position) in the Coulomb field of the nucleus:
$$ H\psi=E\psi,\eqno(1{\rm a})$$
where $H$ is the Hamiltonian
$$
H=c{\bbox\alpha}\cdot{\bf p}+\gamma^0\mu c^2-{\kappa\over r},
\eqno(1{\rm b})$$
where the last term is the Coulomb potential between the electron
(positron) and the nucleus, $\kappa=\pm Ze^2$ where the plus (minus)
sign corresponds to an electron (positron), {\bf p} is the momentum
of the incident electron (positron), ${\bbox\alpha}=\gamma^0{\bbox
\gamma}$, and $\gamma^\mu=(\gamma^0, {\bbox\gamma})$ ($\mu=0,1,2$)
are Dirac matrices satisfying $\{\gamma^\mu, \gamma^\nu\}=2g^{\mu\nu}$
where $g^{\mu\nu}={\rm diag}(1,-1,-1)$. In two dimensions the Dirac
matrices can be realized by $2\times 2$ matrices:
\addtocounter{equation}{1}
\bb
\gamma^0=\sigma^3,\quad \gamma^1=i\sigma^1,\quad
\gamma^2=i\sigma^2,
\ee     %2
where the $\sigma$'s are Pauli matrices.
Thus $\psi$ is a two-component spinorial wave function.
In two dimensions the orbital angular momentum has only one component
\bb
L=\epsilon^{ij}x^ip^j=-i\hbar{\partial\over\partial\theta},
\ee     %3
where $\epsilon^{ij}$ is antisymmetric in its indices and
$\epsilon^{12}=1$, $\theta$ and $r$ used above constitute the polar
coordinates ($r,\theta$) on the $xy$ plane. It is easy to show that
$[L, H]\ne 0$ even for a free particle. However, one can define
another operator
\bb
S={i\hbar\over 4}\epsilon^{ij}\gamma^i\gamma^j,
\ee     %4
and show that
\bb
J=L+S
\ee     %5
is a conserved operator, that is, $[J, H]=0$. Thus $S$ may be
regarded as the spin operator and $J$ the total angular momentum.
$J$ is conserved in any central field.

The Dirac equation (1) can be solved in the polar coordinates by
separation of variables. Bound-state solutions (for $\kappa=Ze^2$)
have been studied in detail [24,25]. We are now considering scattering
solutions with $E>\mu c^2$. We use the representation (2) in which
$S=\hbar\sigma^3/2$. Let
\bb
\psi_j(r,\theta)=\left(
\begin{array}{c}
f(r) e^{i(j-1/2)\theta}/\sqrt r\\
g(r) e^{i(j+1/2)\theta}/\sqrt r
\end{array}
\right),\quad
j=\pm\frac12, \pm\frac32, \ldots.
\ee     %6
It is easy to show that $J\psi_j=j\hbar\psi_j$. Thus $j$ is a good
quantum number. Substituting this expression into Eq. (1) we obtain
two coupled ordinary differential equations for the radial wave
functions:
$$
{df\over dr}-{j\over r}f+k_1g+{\gamma\over r}g=0,
\eqno(7{\rm a})$$
$$
{dg\over dr}+{j\over r}g-k_2f-{\gamma\over r}f=0,
\eqno(7{\rm b})$$
where
\addtocounter{equation}{1}
\bb
k_1={E+\mu c^2\over\hbar c},\quad
k_2={E-\mu c^2\over\hbar c},\quad
\gamma={\kappa\over\hbar c}=\pm{Ze^2\over\hbar c}.
\ee     %8
Then we introduce the new variable
\bb
\rho=kr,\quad k=\sqrt{k_1k_2}={\sqrt{E^2-\mu^2 c^4}\over\hbar c},
\ee     %9
and two new functions $u(\rho)$, $v(\rho)$ through the definition
\bb
f(r)=\frac 12 e^{i\rho}[u(\rho)+v(\rho)],\quad
g(r)=-\frac i2\sqrt{k_2\over k_1} e^{i\rho}[u(\rho)-v(\rho)],
\ee     %10
to recast Eq. (7) into the form
$$
{du\over d\rho}-{i\beta\over \rho}u -{i\beta'+j\over \rho}v=0,
\eqno(11{\rm a})$$
$$
{dv\over d\rho}+2iv+{i\beta\over \rho}v +{i\beta'-j\over \rho}u=0,
\eqno(11{\rm b})$$
where
$$
\beta={\gamma\over 2}\left(\sqrt{k_1\over k_2}+\sqrt{k_2\over k_1}
\right)={\kappa\over\hbar v_{\rm c}},
\eqno(12{\rm a})$$
$$
\beta'={\gamma\over 2}\left(\sqrt{k_1\over k_2}-\sqrt{k_2\over k_1}
\right)=\beta\sqrt{1-{v_{\rm c}^2\over c^2}},
\eqno(12{\rm b})$$
where $v_{\rm c}$ is the classical velocity of the incident particle.
This is more convenient. Indeed, one can eliminate $v$ immediately to
obtain an equation for $u$ alone:
\addtocounter{equation}{2}
\bb
\rho{d^2u\over d\rho^2}+(1+2i\rho){du\over d\rho}+\left(2\beta-
{j^2-\gamma^2\over \rho}\right)u=0,
\ee     %13
where we have used $\beta^2-\beta'^2=\gamma^2$. From Eq. (8),
$\gamma\approx\pm Z/137$. If $Z$ is not very large, say, $Z\le 68$,
we have $|\gamma|<\frac 12$. Then for any $j$, the solution of Eq.
(13) is well behaved at  the origin. Let
\bb
u(\rho)=\rho^s w(\rho),\quad s=\sqrt{j^2-\gamma^2}.
\ee     %14
Then we have for $w$ the equation
\bb
\rho{d^2w\over d\rho^2}+(2s+1+2i\rho){dw\over d\rho}+2(\beta+is)w=0.
\ee     %15
This is familiar. The solution that is well behaved at the origin
is $w(\rho)=\Phi(s-i\beta, 2s+1, -2i\rho)$, where $\Phi(a,b,z)$
is the confluent hypergeometric function. So we have
$$
u_j(\rho)=a_j\rho^s\Phi(s-i\beta, 2s+1, -2i\rho),
\eqno(16{\rm a})$$
where $a_j$ is a constant, and the subscript $j$ of  $u_j$ that is
omitted above has been recovered. Substituting this solution into Eq.
(11a) we have
$$
v_j(\rho)=a_j{s-i\beta\over j+i\beta'}\rho^s
\Phi(s-i\beta+1, 2s+1, -2i\rho),
\eqno(16{\rm b})$$
where we have used the formula
$$
\left(z{d\over dz}+a\right)\Phi(a,b,z)=a\Phi(a+1,b,z),$$
which can be obtained from other relations given in mathematical
handbooks [26]. It should be remarked that $\psi_j$ is slightly
singular at the origin when $j=\pm\frac12$. However, the integral
of $\psi_j^\dagger\psi_j$ over any finite volumn converges and the
solution is acceptable. We take
$$
a_j=A2^s(j+i\beta'){\Gamma(s-i\beta)\over \Gamma(2s+1)}
\exp\left({\beta\pi\over 2}+im\pi-i{s\pi\over 2}+i{\pi\over 4}
\right),
\eqno(17{\rm a})$$
where $m=j-\frac12$ and
$$
A=i\sqrt{E+\mu c^2\over 2E}\sqrt{2\over\pi k},
\eqno(17{\rm b})$$
then we have for the radial wave functions the asymptotic forms
$$
f_j(r)\to Ai^m\exp(i\eta_j)\cos\left(kr+\beta\ln 2kr-{m\pi\over2}
-{\pi\over 4}+\eta_j\right),\quad r\to\infty,
\eqno(18{\rm a})$$
$$
g_j(r)\to\sqrt{k_2\over k_1} Ai^m\exp(i\eta_j)\sin\left(kr+\beta\ln
2kr-{m\pi\over2}-{\pi\over 4}+\eta_j\right),\quad r\to\infty,
\eqno(18{\rm b})$$
to the lowest order, where $\eta_j$ are phase shifts defined through
\addtocounter{equation}{3}
\bb
\exp(2i\eta_j)=(j+i\beta')
{\Gamma(s-i\beta)\over \Gamma(s+i\beta+1)}
\exp(ij\pi-is\pi).
\ee     %19
The asymptotic form for $\psi=\sum_j\psi_j$, where the summation is
taken over all $j$, turns out to be
\bb
\psi\to\psi_{\rm in}+\psi_{\rm sc},\quad r\to\infty,
\ee     %20
where
$$
\psi_{\rm in}=\left(
\begin{array}{c}
i\sqrt{\displaystyle{E+\mu c^2\over 2E}}\\
\sqrt{\displaystyle{E-\mu c^2\over 2E}}
\end{array}
\right)\varphi_{\rm in},
\eqno(21{\rm a})$$
with
$$
\varphi_{\rm in}=\sum_{m=-\infty}^{+\infty}
i^m\sqrt{2\over \pi kr}\cos\left(kr+\beta\ln 2kr-{m\pi\over2}
-{\pi\over 4}\right)e^{im\theta},
\eqno(21{\rm b})$$
and
\addtocounter{equation}{1}
\bb
\psi_{\rm sc}=\sqrt{\frac ir}\exp(ikr+i\beta\ln 2kr)f(\theta)
\left(
\begin{array}{c}
i\sqrt{\displaystyle{E+\mu c^2\over 2E}}\\
e^{i\theta}\sqrt{\displaystyle{E-\mu c^2\over 2E}}
\end{array}
\right),
\ee     %22
with
\bb
f(\theta)=\sqrt{2\over\pi k}\sum_j\exp(i\eta_j)\sin\eta_je^{im\theta}
=-{i\over\sqrt{2\pi k}}\sum_j[\exp(2i\eta_j)-1]e^{im\theta}.
\ee     %23
In order that $\psi$ represents a correct scattering solution, one
should show that $\psi_{\rm in}$ represents an incident wave and
$\psi_{\rm sc}$ represents a scattered one. First, it is easy to
realize that $\varphi_{\rm in}$ is equal to $\exp[i(kr+\beta\ln 2kr)
\cos\theta]$ (for very large $r$).
Thus $\varphi_{\rm in}^*\varphi_{\rm in}=1$. The probability
current density associated with a solution $\psi$ is
\bb
{\bf j}=c\psi^\dagger{\bbox\alpha}\psi.
\ee     %24
For $\psi_{\rm in}$, it is easy to show that
\bb
{\bf j}_{\rm in}=v_{\rm c}{\bf e}_x,
\ee     %25
where ${\bf e}_x$ is the unit vector in the $x$ direction. For
$\psi_{\rm sc}$ we have
\bb
{\bf j}_{\rm sc}={|f(\theta)|^2\over r}v_{\rm c}{\bf e}_r,
\ee     %26
where ${\bf e}_r$ is the unit vector in the radial direction. Thus
$\psi$ is indeed a correct scattering solution, $f(\theta)$ is the
scattering amplitude, and the differential cross section is given by
\bb
\sigma(\theta)=|f(\theta)|^2.
\ee            %27
The choice of $a_j$ in Eq. (17) is thereby proved to be appropriate.
The results (23) and (27) have the same forms as those for a
short-range central field [27], but here the asymptotic forms (18)
where the phase shifts occur involve the logarithmic distortion which
is typical for a Coulomb field. Since $\sum_j e^{im\theta}=2\pi\delta
(\theta)$, we have for $\theta\ne 0$
\bb
f(\theta)
=-{i\over\sqrt{2\pi k}}\sum_j\exp(2i\eta_j)e^{im\theta}.
\ee     %28
This result with $\exp(2i\eta_j)$ given by Eq. (19) is exact. It is
similar to but simpler than the corresponding result in three
dimensions. However, it is very difficult to sum up  the
above partial wave series in the general case. Closed results can
be obtained approximately only for light nucleus (small $Z$).

Consider light nucleuses which may be more interesting in pratice.
With $\gamma\approx Z/137$ and, say, $Z<5$, we have $\gamma^2\ll 1$.
Then we may approximately replace $s$ by $|j|$ in Eq. (19). Note that
$\beta$ also depends on $\gamma$ and we do not make approximation with
it. Thus the result will possess some nonperturbative features. We
have
$$
\exp(2i\eta_j)={\Gamma(m+1/2-i\beta)\over \Gamma(m+1/2+i\beta)}
-i(\beta-\beta'){\Gamma(m+1/2-i\beta)\over \Gamma(m+3/2+i\beta)},
\quad (j>0),
\eqno(29{\rm a})$$
$$
\exp(2i\eta_j)={\Gamma(|m|+1/2-i\beta)\over \Gamma(|m|+1/2+i\beta)}
+i(\beta-\beta'){\Gamma(|m|-1/2-i\beta)\over \Gamma(|m|+1/2+i\beta)},
\quad (j<0).
\eqno(29{\rm b})$$
It can be shown that the first term in either equation equals
$\exp(2i\delta_{|m|})$ where $\delta_{|m|}$ is the nonrelativistic
phase shift of the $m$th  partial wave. The second term is a
relativistic correction which vanishes when $v_{\rm c}/c\to 0$.
Substituting Eq. (29) into Eq. (28) we have
\addtocounter{equation}{1}
\bb
f(\theta)=f_0(\theta)+f_1(\theta),
\ee     %30
where
\begin{eqnarray}
f_0(\theta)=-{i\over\sqrt{2\pi k}}&&\left[
{\Gamma(1/2-i\beta)\over \Gamma(1/2+i\beta)}F(1,1/2-i\beta,
1/2+i\beta, e^{i\theta})\right. \nonumber \\
+&&\left.{\Gamma(3/2-i\beta)\over \Gamma(3/2+i\beta)}e^{-i\theta}
F(1,3/2-i\beta,3/2+i\beta, e^{-i\theta})\right]
\end{eqnarray}     %31
is the nonrelativistic partial wave result and
\bb
f_1(\theta)=-{\beta-\beta'\over\sqrt{2\pi k}}
{\Gamma(1/2-i\beta)\over \Gamma(3/2+i\beta)}[F(1,1/2-i\beta,
3/2+i\beta, e^{i\theta})-e^{-i\theta}
F(1,1/2-i\beta,3/2+i\beta, e^{-i\theta})]
\ee     %32
is the relativistic correction. In these equations
$F(a_1, a_2, b_1, z)$ is the hypergeometric function. Using their
functional relations (cf Eqs. (e.6) and (e.4) in Ref. [2]),
it can be shown that
\bb
f_0(\theta)=-i{\Gamma(1/2-i\beta)\over \Gamma(i\beta)}
{\exp(i\beta\ln\sin^2\theta/2)\over
\sqrt{2k}\sin\theta/2},
\ee     %33
and
\bb
f_1(\theta)=-{\Gamma(1/2-i\beta)\over \Gamma(i\beta)}
\left(1-{\beta'\over\beta}\right){e^{-i\theta/2}
\exp(i\beta\ln\sin^2\theta/2)\over\sqrt{2k}},
\ee     %34
where $0\le\theta<2\pi$. It should be remarked that both terms in
Eq. (31) diverge for all $\theta$ [26]. The final result (33) is well
defined everywhere except at $\theta=0$, however. Therefore the
calculation that leads to Eq. (33) may also be regarded as a
regularization procedure. The validity of this procedure is obvious
because the final result (33) coincides with the nonrelativistic
one  obtained from exact solutions [23]. On the other hand, both terms
in Eq. (32) converge everywhere except at $\theta=0$ [26]. Thus the
calculation of $f_1(\theta)$ involves nothing special. Collecting the
above results we arrive at 
\bb
f(\theta)=-i{\Gamma(1/2-i\beta)\over \Gamma(i\beta)}
{\exp(i\beta\ln\sin^2\theta/2)\over
\sqrt{2k}\sin\theta/2}\left[1-ie^{-i\theta/2}\sin
{\theta\over 2}\left(1-{\beta'\over\beta}\right)\right].
\ee     %35
We have dropped a $\delta(\theta)$ term in Eq. (28), since it does
not contribute when $\theta\ne 0$. Here we have a result more singular
than $\delta(\theta)$ when $\theta\to 0$. Thus the $\delta(\theta)$
term can indeed be dropped everywhere and the above expression is
enough. It is easy to find that
\bb
\sigma(\theta)=|f(\theta)|^2
={\beta\tanh\beta\pi\over 2k\sin^2\theta/2}
\left(1-{v_{\rm c}^2\over c^2}\sin^2{\theta\over 2}\right).
\ee            %36
Two remarks about the result: First, we have not made any
approximation in regard to the incident velocity, so the result is
valid for high energy collision. It is obvious that the relativistic
correction becomes significant when $v_{\rm c}$
is comparable with $c$.
Second, though the above result holds for small $\gamma$ only, it
involves a nonperturbative factor $\tanh\beta\pi$ (note that
$\beta\propto\gamma$). Thus the result cannot be obtained from
perturbative QED at the tree level. This is different from the case
in three dimensions where the result corresponding to Eq. (36)
coincides with the lowest-order contribution of perturbative QED.
The reason may be that in three dimensions the Born approximation
(a perturbative result) already gives the correct result in the
nonrelativistic  limit. In two dimensions this is not true.
If one expands $\exp(2i\eta_j)$ to a higher order in $\gamma$ by
expanding $s$ and if these higher-order corrections can be summed
up analytically to give a higher-order correction $f_2(\theta)$ to
$f(\theta)$, then it might be expected that  $f_2(\theta)$ also
possesses the above nonperturbative feature since no approximation
is made in regard to $\beta$. The higher-order corrections will be
studied subsequently.
Finally we express $\sigma(\theta)$ in terms of pure classical
quantities:
$$
\sigma(\theta)
={\kappa\tanh(\pi\kappa/\hbar v_{\rm c})\over 2\mu v_{\rm c}^2
\sin^2\theta/2}
\left(1-{v_{\rm c}^2\over c^2}\sin^2{\theta\over 2}\right)
\left(1-{v_{\rm c}^2\over c^2}\right)^{\frac 12}.
\eqno(36')$$
Here the first factor is the exact nonrelativistic result. The
subsequent ones are due to the relativistic effect. They are very
similar to those in three dimensions.

In conclusion, we have calculated the scattering amplitude for fast
electrons (positrons) scattered by a nucleus in two dimensions.
The exact result is given by Eq. (23) in partial wave series. For
light nucleuses, the series can be summed up approximately to give a
closed result (35). The differential cross section is given by
Eq. (36) or ($36'$). Though being approximate, the results exhibit
some nonperturbative feature and cannot be obtained from the
lowest-order contribution of perturbative QED. This feature is not
manifest in  the corresponding result in three dimensions.

\vskip 1pc

The author is grateful to Professor Guang-jiong Ni for encouragement.
This work was supported by the
National Natural Science Foundation of China.

%\newpage

\end{document}